\documentclass[prl,twocolumn,superscriptaddress,floats]{revtex4}

\usepackage[dvips]{graphicx}

\usepackage{amsmath}

\usepackage{dcolumn}

\usepackage{rotating}

\newlength{\La} 
\newlength{\Lb} 
\newlength{\Lc} 

\newcolumntype{d}{D{.}{.}{-1}}

\newcommand{\bkbo}{BKBO}

\newcommand{\lsco}{La$_{2-x}$Sr$_{x}$CuO$_4$}

\begin{document}

\advance\vsize by 2 cm
\title{Giant phonon anomalies in the bond-stretching branches in {Ba$_{0.6}$K$_{0.4}$BiO$_3$} }

\author{M.~ Braden}
\affiliation{Forschungszentrum Karlsruhe, IFP, Postfach 3640, D-76021
Karlsruhe, Germany}
\affiliation{Laboratoire L\'eon Brillouin,
C.E.A./C.N.R.S., F-91191-Gif-sur-Yvette CEDEX,
France}

\author{W. Reichardt}
\affiliation{Forschungszentrum Karlsruhe, IFP, Postfach 3640, D-76021
Karlsruhe, Germany}

\author{S. Shiryaev}
\affiliation{Institute of Solids and Semiconductors, Acad. Sci., P. Brovski 17,
220072 Minks, Belarus}

\author{S.N. Barilo}
\affiliation{Institute of Solids and Semiconductors, Acad. Sci., P. Brovski 17,
220072 Minks, Belarus}

\date{\today, \textbf{DRAFT}}

\pacs{PACS numbers:}

\begin{abstract}

The dispersion of the longitudinal bond-stretching (BS) branches
in superconducting {Ba$_{0.6}$K$_{0.4}$BiO$_3$} (BKBO) ~
studied by inelastic neutron scattering reveals giant phonon anomalies.
In the [110] and [111] directions these branches even split close to
the middle of the zone, in contradiction to any harmonic
lattice dynamical model with cubic symmetry. 
The additional low energy branch
in the [111] direction ends at the {\bf R}-point with a frequency
near 10.5THz representing a doping induced renormalization of
a breathing type vibration by about 40\%, the strongest observed so far in any
superconductor.

\end{abstract}

\maketitle

\bkbo ~ merits special interest since it exhibits by far the
highest T$_c$ among the non-cuprate perovskite superconductors
\cite{cava,matheis}.
The pure compound exhibits charge order
between Bi$^{3+}$ and Bi$^{5+}$ \cite{cox} associated with a 
breathing distortion of the lattice, 
i.e. alternating BiO$_6$-octahedra with short
(Bi$^{5+}$) and long (Bi$^{3+}$) Bi-O-bond distances.
The Bi-O bond distances reflect directly the charge in the
octahedron.
In consequence the BS vibrations are
strongly coupled to any form of charge inhomogeneity.
The mode at  {\bf R}=(0.5 0.5 0.5), 
exactly reflects the breathing distortion and is called the breathing mode.
The zone-boundary modes at {\bf X}=(0.5 0 0) and {\bf M}=(0.5 0.5 0),
called linear and planar breathing modes, are depicted in the insets 
in figures 1 and 2.

Our previous phonon
studies on superconducting \bkbo ~with x$\sim$0.37
revealed doping induced renormalization of the BS modes
in all directions :
a strong bending down of the 
branch in the [100]-direction, but flatter dispersion
along [110] and [111] \cite{bkbo-epl,bkbo-jos}
which did not agree with the predicted charge density wave 
instabilities just for the latter directions \cite{hahn}.
Phonon anomalies in the BS branches have also been found in 
other perovskites \cite{pini-reich,reich-bra,lanio}; 
it seems to be a general property
of these metallic oxides
that a close charge order instability 
\cite{tranquada,review}
is reflected in a strong
renormalization of BS modes.



Due to the recent progress in the crystal growth of \bkbo ~ \cite{barilo}
(almost a factor of ten gain in sample volume \cite{bkbo-str}) more detailed
studies of the phonon dispersion can now be performed.
These studies reveal 
anomalies even more complex than previously observed; 
in particular the new results show that phonon anomalies 
are much more pronounced in \bkbo ~ than
in superconducting cuprates.

Two of the large crystals described in reference \cite{bkbo-str}
with almost identical K-contents of x=0.40 and 0.41 were used in this work.
Inelastic neutron scattering was performed on the triple-axis 
spectrometer 1Ta using a Cu-(111) monochromator
and a  PG-(002) analyzer.
The final energy was fixed either to
3.55THz or 7.37THz; the latter configuration was used 
in most cases in order to avoid 
higher order contamination.
Samples were cooled to $\sim$11K
to minimize background;
only a few comparative scans were performed at higher temperature.

Along the [100]-direction the BS branch connects to the
linear breathing mode at the zone boundary, {\bf X}.
In figure 1 we show scans on these modes for {\bf q}=(0 0 0) to {\bf X},
{\bf Q}=(5 0 0) to (4.5 0 0) ({\bf Q} denotes the scattering vector with 
{\bf Q}={\bf q}+${\bf \tau}$ where {\bf q} is within the first Brillouin zone and 
${\bf \tau}$ a reciprocal lattice vector).
In agreement
with our previous report, the branch 
shoots down in frequency near  {\bf q}=(0.25 0 0), see Fig. 1.
In addition to the frequency softening along [100]
the modes are strongly broadened close to {\bf X}.
The improved statistics also permits a quantitative analysis
of the intensity : the upper part of figure 1
clearly demonstrates, that the
integrated intensity continuously decreases when approaching 
{\bf X} in the range (0.3 0 0) to (0.5 0 0), i.e. in the range where
frequency is almost constant. The fact that this behavior is seen
in two configurations (and also for other {\bf Q}-values not shown)
excludes an experimental artifact.
The loss of intensity towards {\bf X} may suggest that some spectral weight
is split off and shifted to even lower energy.

\begin{figure}[t]
\begin{center}
\includegraphics*[width=0.84\columnwidth]{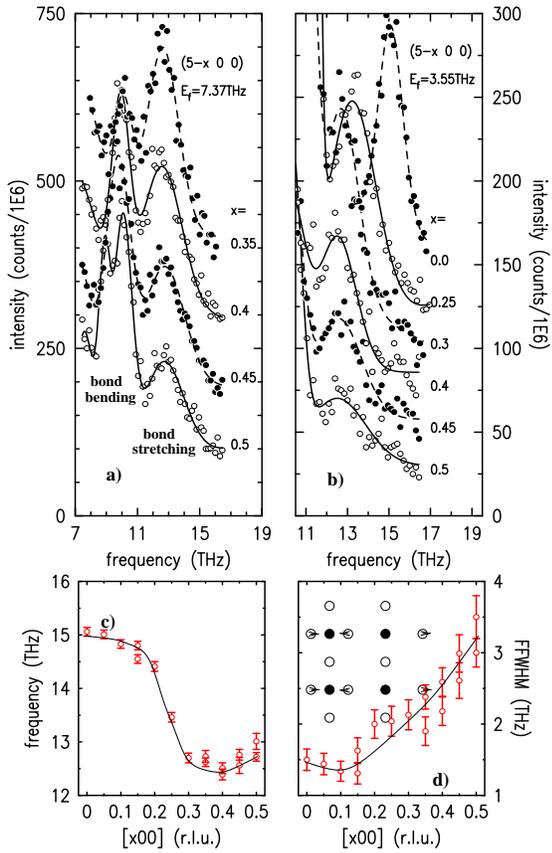}%
\caption{Scans on the BS modes in 
the [x00] direction at {\bf Q}=(5-x 0 0)
with low a) and high b) resolution.
Dispersion of the BS mode along [100], c), and {\bf q}-dependence
of the FWHM, d); the inset in d) shows an elongation pattern
of the 1D-breathing mode at {\bf q}=(0.5 0 0) for a single BiO$_2$-layer;
closed circles denote Bi- and open ones O-ions.
} \label{fig:1}
\end{center}
\end{figure}

Figure 2 and 3 present the scans recorded on the BS modes
in the [110] and [111] directions.
Along [110] the scans were obtained by comparing and
adding the results at many different {\bf Q}-values.
Along [111], all scans were performed at {\bf Q}=(4-x 1+x 1+x) which allows 
direct comparison of intensity.
Additional measurements were also done along [111] 
in many different Brillouin zones
in order to establish the dispersion unambiguously.
In both directions we find an unexpected splitting :
upon increase of x in {\bf q}=(xx0) and in {\bf q}=(xxx)
the unique peak at $\Gamma$ develops first a shoulder  for
x=0.1--0.2 and then shows intensity clearly separated in energy
up to the zone boundary.
The lower features were not
detected in the first study
due to a lower signal to background ratio and
to much lower counting statistics. Furthermore, the previous measurements
performed with E$_f$=3.55THz could not be extended to low energies
due to an intrinsic contamination at 10.5\ THz.
At the low energy side of the scans in figures 2 and 3 one detects the
longitudinal bond-bending (BB) modes which dominate the spectral distribution,
in agreement with the lattice dynamical model calculations.
The interpretation that the split off intensities
are of BS character is further corroborated by intensity
checks. When comparing the total intensity of the split peaks with that
of the BS mode at $\Gamma$ and with those of the 
BB modes at {\bf M} and {\bf R}, respectively, we find good agreement
with calculated structure factors by a lattice dynamical model, whereas 
the upper better-defined peaks account only for about
half of the expected intensity.

Figure 4 shows the dispersion of the BS branches in
[110] and [111] directions deduced from a two peak analysis
of the experimental spectra. The upper branches agree well with
our former results \cite{bkbo-epl,bkbo-jos}. 
In the [110]-direction the additional
branch is close to the low frequency plateau
of the [100] branch. In the [111] direction the
lower branch exhibits continuous softening through the zone, ending
near 10.5THz at {\bf R}. 
However, it cannot be excluded that even the additional branch is 
split, with an upper branch with reduced slope and
a dispersion-less feature near 10.5\ THz. 
The interpretation of the additional spectral weight is severely 
hampered by the strong intensities from the
LO branches with BB character. Whereas in the [111]
direction this branch stays below 9 THz the situation is 
much less favorable for the [100] and [110] directions.
Here the corresponding branches rise from below to beyond 10 THz 
near the zone boundaries. For this reason we cannot exclude
that near {\bf X} and {\bf M} some of the intensity in the 10 THz region
has to be attributed to BS vibrations. 
It appears interesting to note that in the closely related compound
BaBi$_{0.25}$Pb$_{0.75}$O$_3$
a dispersion-less additional intensity has been found 
just at this energy \cite{bpbo}.

\begin{figure}[t]
\begin{center}
\includegraphics*[width=0.8\columnwidth]{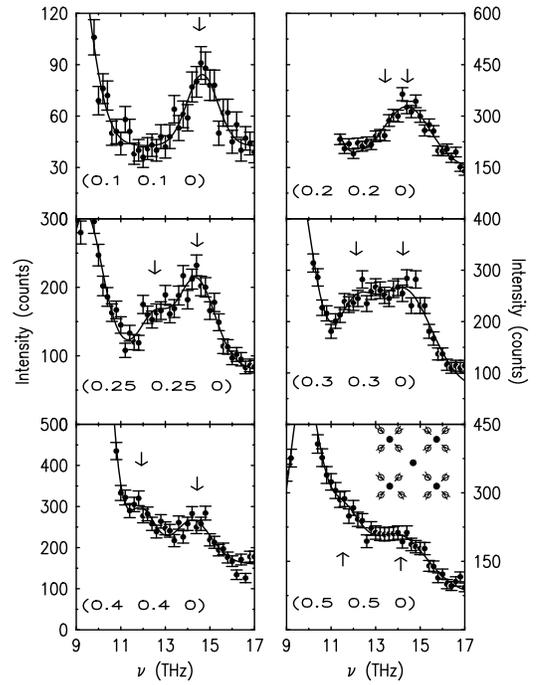}%
\caption{
Scans on the BS modes in [110] direction, scans at different
equivalent Q-values have been added, E$_f$=7.37THz; 
arrows indicate the positions of the modes.
The inset shows an elongation pattern
of the  planar breathing mode at {\bf q}=(0.5 0.5 0) for a single BiO$_2$-layer;
closed circles denote Bi- and open ones O-ions.
} \label{fig:2}
\end{center}
\end{figure}

Two groups have examined the phonon density of states upon doping
\cite{loong,prassides}; there is a strong
doping induced shift of spectral weight from the region 
above 15\ THz towards
the region 10--13THz. This behavior agrees qualitatively with the bending down
of the BS branch along [100] and the additional
low frequency intensities displayed in figure 4. 
The simulation of the phonon density of states
with the lattice dynamical model indicates that the observed
spectral shift cannot be explained by the bending 
down of the [100]-branch 
alone, which provides additional support for the
identification of the split intensities to be 
of BS character.

The temperature dependence of the BS intensities
has been studied at (0.4 0 0), {\bf X} and {\bf R} on the crystal
with the superconducting transition at T$_c$=29.5\ K.
We observe no difference in position or peak width between
10, 36 and 120\ K, whereas the phonons become significantly broadened
upon heating to room temperature. The absence of an anomaly at T$_c$
does not contradict the expected strong electron phonon coupling,
since the phonon frequencies are much higher than the superconducting gap.

\begin{figure}[t]
\begin{center}
\includegraphics*[width=0.8\columnwidth]{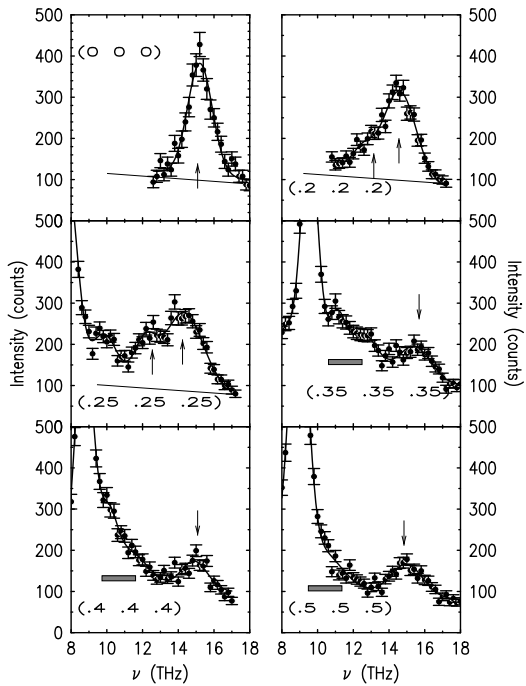}%
\caption{
Scans on the BS modes in [111] direction
at {\bf Q}=(4-x 1+x 1+x), E$_f$=7.37THz;
arrows or horizontal bars indicate the positions of additional intensity;
thin lines show the background .
} \label{fig:3}
\end{center}
\end{figure}

Obviously, the two branches in figure 4 cannot be explained
by any model based on
harmonic lattice dynamics and ideal cubic perovskite symmetry.
The split dispersion might arise from a folding back of the zone
boundary frequencies to the $\Gamma$-point.
In principle the rotation distortion present
in these crystals \cite{bkbo-str} might cause such behavior;
however, lattice dynamical calculations within the symmetry
of this distortion show that there is only minor influence
of the rotation on the BS modes, and,  in particular,
there are no gaps at the zone-boundary of the superstructure.
Since Ba/K-mixing may in principle cause splittings in the phonon dispersion,
we have analyzed these effects by superlattice calculations with the
Ba and K randomly distributed;  details of these calculations are
described in Ref. \cite{bkbo-str}. The simulated spectra yield
an influence for the BB branches, but there is no sizeable effect
on the BS branches; this finding reflects the orientation
of the Ba/K-O interaction perpendicular to the Bi-O-bonds.


\begin{figure}[t]
\begin{center}
\includegraphics*[width=0.8\columnwidth]{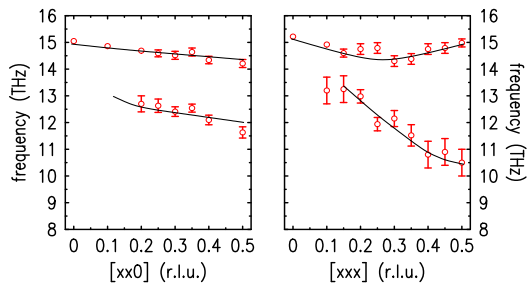}%
\caption{Dispersion of the split longitudinal BS branches
along the [110] and [111]-directions, lines are guides to the eye.
} \label{fig:4}
\end{center}
\end{figure}

The renormalization of the BS frequencies
can be attributed to electron lattice interaction.
LDA band structure calculations indeed predicted
some frequency softening induced by doping 
\cite{band-str,band-str3};
however, the agreement is at most qualitative
and the splitting of the BS branches is not
explained at all.
The observed much stronger effects indicate that ab initio
LDA calculation \cite{band-str3} severely underestimates the electron phonon
coupling for the BS modes.
The strength of
the   renormalization and the splitting in the BS branches 
must be directly related to the
bond distances and, hence, to some inhomogeneity or fluctuation of charge.
For example, the three-dimensional breathing mode at {\bf R} is associated
with oscillating octahedral volumes. 
If the charge in the octahedra
is oscillating in phase  -- i.e. positive charge is always
going to the smaller octahedron --,
the phonon frequency becomes reduced similar to the reduction of a 
longitudinal polar frequency by screening.
One may consider specific charge fluctuations
to screen the linear and planar breathing modes too.
However, such screening of BS displacements 
requires at least a partial localization of the charges on
a time scale determined by the inverse phonon frequency.
Dynamic charge-phonon coupling may explain a continuous 
dispersion along [100], but, the split dispersion suggests
some charge inhomogeneity on a time scale larger than the inverse
phonon frequency.
Evidence for some local variation of the Bi-O-bond distance,
and hence charge inhomogeneity,
has been found in EXAFS-experiments \cite{exafs} and 
in the anisotropic Debye-Waller factors \cite{bkbo-str}.
One should emphasize the strong frequency broadening of all renormalized
modes. The FWHM of the mode at {\bf X} indicates a phonon life-time
of the order of the vibration period; the even broader signals in the
split-of branches along [110] and [111] demonstrate that the character 
of these excitations
is quite different to that of a harmonic phonon mode.
One may speculate, that also the spatial extension of these
excitations is reduced.

The observations in \bkbo ~ may be compared 
with other metallic perovskites.
The shooting down of the BS branch in [100]
is observed in manganates \cite{reich-bra}, 
in  both cuprate systems studied, La$_{1.85}$Sr$_{0.15}$CuO$_4$ 
and YBa$_2$Cu$_3$O$_{7-\delta}$ \cite{pini-reich}, 
and in nickelates \cite{lanio}. 
In manganates, cuprates and the bismuthates the step-like form
of the dispersion is extremely anomalous, and the similarity between these
systems is striking. 
The step near {\bf q}=(0.25 0 0) suggests an underlying 
cell-doubling. McQueeny et al. \cite{mcqueeny}
have tried to explain the dispersion in
La$_{1.85}$Sr$_{0.15}$CuO$_4$ by such a superstructure; 
however, recent studies did not reveal the expected 
discontinuity in the dispersion \cite{pini-lsc}.
We think that the step-like shooting down in all these perovskites
arises from a dynamic charge lattice coupling. 
The [100] BS modes appear to be exceptional due to their
particular displacement pattern : 
the metal-ions with close oxygens are  connected
via metal-oxygen-metal bonds, see figure 1.
This displacement pattern further yields a pronounced coupling to the 
charge on the oxygen.
Furthermore, 
this configuration requires only little reduction of the kinetic 
energy of the charges involved in the screening, 
since the metal-oxygen-metal paths will favor metallic behavior
even in a frozen-in phase.
For the displacement patterns of the BS branches
along the other directions, the cations with short bonds are
not connected by metal-oxygen paths; therefore, the screening of the
displacements by some charge displacement requires sizeable reduction 
of kinetic energy.

The outstanding renormalization along 
[110] and [111] found in the bismuthate does not have an 
analogy in the superconducting cuprates \cite{pini-reich}.
One may note that the spectral weight of the 
frequency shifts is essentially enhanced
by the additional renormalization along [110] and [111], since these directions
show high multiplicities of 12 and 8 respectively. Therefore, 
the spectral weight of the BS mode renormalization in 
\bkbo , which may be taken as a measure of the electron phonon coupling,
is essentially larger than that found in the cuprates,
rendering a superconducting mechanism based on coupling to 
BS modes \cite{weber} much more likely for \bkbo .
The distinct behavior along [110] and [111]  
between manganates and bismuthates on one side 
and the cuprates on the other side, is further reflected
in different charge-ordering schemes appearing upon substitution
\cite{review,hahn,cox};
cuprates exhibit only one-dimensional stripe-type order \cite{tranquada}.
Finally, we note that a split dispersion 
is observed for the first time in  \bkbo , but 
should represent a characteristic feature for perovskites with 
inhomogeneous charge distribution.

In conclusion, we have reported on the dispersion of the
BS vibrations in \bkbo ~ which are strongly renormalized
in comparison to BaBiO$_3$.
We confirm the sharp shooting down of the phonon frequency
in the [100]-direction observed previously, 
which closely resembles the observation
in manganates and superconducting cuprates.
In addition, in \bkbo ,  the branches along [110] and [111] split, with
the lower frequency reaching 10.5THz at {\bf R}.
Compared to undoped insulating BaBiO$_3$, where the breathing
mode frequency amounts to 17\ THz,  this represents a 
doping-induced frequency renormalization
of about 40\%.
The renormalised modes exhibit outstanding 
frequency broadening.
All these observations
suggest that phonons are strongly coupled to
some charge fluctuations  in \bkbo .
Comparing \bkbo ~ to  \lsco ~ and YBa$_2$Cu$_3$O$_7$, 
which exhibit similar or higher superconducting transition
temperatures, the spectral shift due to the BS phonon anomalies
in \bkbo ~ is substantially larger.

Part of this work was supported through INTAS by project 1371.

\end{document}